\documentclass[a4paper,11pt]{article}
\usepackage{jheppub} 
\usepackage{lineno}

\usepackage{graphics}
\usepackage{amsmath}
\usepackage{hyperref}
\usepackage{amssymb}
\usepackage{psfrag}
\usepackage{epsfig}
\usepackage{epsf}
\usepackage{float}
\usepackage{slashed}
\usepackage{dsfont}
\usepackage{bbold}
\usepackage{color}
\allowdisplaybreaks

\newcommand{\ben}{\begin{displaymath}}
\newcommand{\een}{\end{displaymath}}
\newcommand{\be}{\begin{equation}}
\newcommand{\ee}{\end{equation}}
\newcommand{\bea}{\begin{eqnarray}}
\newcommand{\eea}{\end{eqnarray}}

\allowdisplaybreaks

\arxivnumber{2312.05193} 

\title{Gravitational $p \to \Delta^+ $ transition form factors in
  chiral perturbation theory}

\author[a]{H.~Alharazin,}
\author[a]{B.-D. Sun,}
\author[a]{E.~Epelbaum,}
\author[a,b]{J.~Gegelia,}
\author[c,d,b]{U.-G.~Mei\ss ner}
 \affiliation[a]{Institut f\"ur Theoretische Physik II, Ruhr-Universit\"at Bochum,  D-44780 Bochum,
 Germany}
 \affiliation[b]{Tbilisi State  University,  0186 Tbilisi,
 Georgia}
 \affiliation[c]{Helmholtz Institut f\"ur Strahlen- und Kernphysik and Bethe
   Center for Theoretical Physics, Universit\"at Bonn, D-53115 Bonn, Germany}
 \affiliation[d]{Institute for Advanced Simulation, Institut f\"ur Kernphysik
   and J\"ulich Center for Hadron Physics, Forschungszentrum J\"ulich, D-52425 J\"ulich,
Germany}

\abstract{The gravitational form factors of the transition from the proton to
the $\Delta^+$ resonance are calculated to leading one-loop order using
a manifestly Lorentz-invariant formulation of chiral perturbation
theory. We take into account the leading electromagnetic and
strong isospin-violating effects. The loop contributions to the
transition form factors are found to be free of power-counting
violating pieces, which is consistent with the absence of
tree-level diagrams at the considered order. In this sense, our
results can be regarded as predictions of chiral perturbation theory. 
}

\begin{document}
\maketitle
\flushbottom

\section{Introduction}

The linear response of the effective action to the change of the space-time metric
specifies mechanical properties of particles.
In particular, static characteristics, like the mass, spin and the $D$-term correspond to the hadron
gravitational form factors (GFFs) at zero momentum transfer \cite{Kobzarev:1962wt,Pagels:1966zza}.
In recent years, GFFs have attracted increasing attention for characterizing properties of
hadrons with different spins due to their connection to generalized parton distributions (GPDs). 
Parameterizations of the energy-momentum tensor (EMT) matrix elements
in terms of the GFFs have been considered
for spin-$0$~\cite{Pagels:1966zza}, spin-1~\cite{Holstein:2006ud,Cosyn:2019aio,Polyakov:2019lbq},
and for arbitrary-spin hadrons~\cite{Cotogno:2019vjb}. 
Mechanical properties, energy and spin densities as well as spatial distributions of
the pressure and shear forces have been introduced for spin-$0$ and spin-$1/2$ hadrons in
Ref.~\cite{Polyakov:2002yz}, 
and generalized to higher-spin systems in Refs.~\cite{Polyakov:2019lbq,Panteleeva:2020ejw,Kim:2020lrs}.

The nucleon GFFs can be extracted from experimental measurements of
exclusive processes like deeply virtual Compton scattering  (DVCS)
\cite{Ji:1996ek,Radyushkin:1997ki} and hard exclusive meson production \cite{Collins:1996fb}.
The connection to GFFs can be seen in the QCD description of these processes, where the
{\it symmetric} EMT   appears naturally in the operator product expansion \cite{Ji:1996ek}.  
The first  results of  measurements of the $D$-term
in hard QCD processes for the nucleon and the pion can be found in
Refs.~\cite{Kumericki:2015lhb,Nature,Kumano:2017lhr,Kumericki:2019ddg}. 
Recently, the mechanical radius of the proton has been determined from
experimental data on DVCS cross sections and polarized electron beam
spin asymmetries \cite{Burkert:2023atx}.
The GFFs  have also been studied in lattice QCD, see, e.g.,
Refs.~\cite{Shanahan:2018nnv,Shanahan:2018pib,Alexandrou:2013joa,Bratt:2010jn,Hagler:2007xi,Hackett:2023rif}
and references therein.

  While the electromagnetic $p \to \Delta^+$ transition has been
  extensively studied over the past two decades on both the theoretical
  and experimental sides, see, e.g.,
  Refs.~\cite{Hilt:2017iup,Blomberg:2015zma,CLAS:2009ces,Tiator:2011pw,Tiator:2016btt},
  the gravitational $p \to \Delta^+$ transition form factors (GTFFs)
  gained attention only since a few years \cite{Kim:2022bwn}.
  The GTFFs can be accessed experimentally through their connection to the transition GPDs \cite{Frankfurt:1998jq,Frankfurt:1999xe},
  obtained by expanding the non-local QCD operators 
  with various quantum numbers. 
Non-perturbative properties of the nucleon-$\Delta$ transition GPDs have been studied, e.g., by applying the approach of large $N_c$ limit of QCD, as discussed in Sec.~2.7 of 
Ref.~\cite{Goeke:2001tz}.  In Ref. \cite{Semenov-Tian-Shansky:2023bsy}, the transition GPDs have been connected with
  the DVCS amplitude within the process $e^- N \to e^- \gamma \pi N,$ while
  in Ref. \cite{Kroll:2022roq} these quantities have been studied using exclusive electroproduction
  of $\pi^- \Delta^{++}$.
     
     In Ref.~\cite{Kim:2022bwn}, the matrix element of the symmetric
     EMT corresponding to the $p \to \Delta^+$  transition has been
     studied for the first time, where a parametrization for the
     transitions $\frac{1}{2}^{\pm} \to \frac{3}{2}^{\pm} $ and
     $\frac{1}{2}^{\pm} \to \frac{3}{2}^{\mp} $ has been suggested in
     terms of five conserved and four non-conserved GTFFs. The first
     calculations of the GTFFs of the $N \to \Delta$ transition were
     done in Ref.~\cite{Ozdem:2022zig} using the QCD light-cone sum
     rules. The interpretation and understanding of the GTFFs have generated much
     interest recently. In particular, the concept of QCD angular
     momentum (AM)~\cite{Leader:2013jra,Lorce:2017wkb,Granados:2019zjw} has been extended to $N \to \Delta$ transitions  in
     Ref.~\cite{Kim:2023xvw}. These quantities were calculated in the
     $1/N_c$  expansion,  and their connection to the transition GPDs
     of the hard exclusive electroproduction processes was
     discussed. Properties of the AM of various transitions were further
     explored in Ref.~\cite{Kim:2023yhp}, where their decomposition into the
     orbital AM and the intrinsic spin components was studied.
 
For systematic studies of  low-energy hadronic processes involving the
$\Delta$ resonances and 
induced by gravity one may rely on the effective chiral Lagrangian for the nucleons, pions,
photons and delta resonances in  curved spacetime.
Effective Lagrangian of pions in curved spacetime has been derived in Ref.~\cite{Donoghue:1991qv}, 
and the GFFs of the pion are considered in Ref.~\cite{Kubis:1999db}. 
The leading and subleading effective chiral Lagrangians for nucleons, delta resonances and
pions in curved spacetime,
along with the calculation of the leading one-loop 
contributions to the GFFs of the nucleons and the $\Delta$ resonances  can be found in
Refs.~\cite{Alharazin:2020yjv,Alharazin:2022wjj}. 

     In this work we calculate the GTFFs of the $p \to \Delta^+$
     transition in the
     framework of manifestly Lorentz-invariant chiral perturbation
     theory (ChPT) up-to-and-including the third order in the small-scale expansion
     \cite{Hemmert:1996xg}. As gravity conserves isospin, such kind of processes are possible
     only if the isospin symmetry is
     broken, i.e.~if $m_u \neq m_d$ and/or if the electromagnetic interaction is taken into
     account. We include both effects at the corresponding leading orders to calculate the
     one-loop contributions to the GTFFs.

Our paper is organized as follows: In section~\ref{I}, we specify the
relevant terms of the  effective Lagrangian of the nucleons, pions, photons and delta resonances in
curved spacetime.   
We calculate the GTFFs of the $p \to \Delta^+$ transition in section~\ref{II}.  
The results of our calculations are summarized in
section~\ref{summary}. In the appendices, we list the isospin symmetry
breaking terms in the action and the expression for the parts of the EMT, which
are relevant for our study.

\section{Effective Lagrangian in curved spacetime and the energy-momentum tensor}
\label{I}

The action corresponding to the leading-order effective Lagrangian for
nucleons, pions, photons and delta resonances, interacting with an
external gravitational field, can be easily obtained from the
corresponding expressions in flat spacetime
\cite{Donoghue:1991qv,Gasser:1987rb,Fettes:2000gb,Hemmert:1997ye,Hacker:2005fh}.
It has the following form:
\begin{eqnarray}
S_{\rm \gamma}^{(2)}  
& = &
\int d^4x \sqrt{-g}\, \biggl\{\, -\frac{1}{4} F_{\mu \nu }F^{\mu \nu } + \frac{m_\gamma^2}{2} A_\mu A^\mu \biggl\},
\label{Gg}
\\
S_{\rm \pi}^{(2)} 
&=& \int d^4x \sqrt{-g}\, \left\{ \frac {F^2}{4}\,  {\rm Tr}
( D_\mu U  (D^\mu U)^\dagger ) + \frac{F^2}{4}\,{\rm Tr}(\chi U^\dagger +U \chi^\dagger) \right\}\,,
\label{PionAction} 
\\
S_{\rm N \pi}^{(1)}  
& = & 
\int d^4x \sqrt{-g}\, \biggl\{\, \bar\Psi \, i  \gamma^\mu
\overset{\leftrightarrow}{\nabla}_\mu \Psi -m \bar\Psi\Psi  +\frac{g_A}{2}\, \bar\Psi \gamma^\mu \gamma_5 u_\mu \Psi  \biggr\} \,, 
\label{PiNAction} 
\\ 
S_{\Delta \pi}^{(1)} 
& = &
 - \int d^4 x  \sqrt{-g} \biggl\{  
\Bar{\Psi}^{i \mu}  \,  i \gamma^\alpha \overset{\leftrightarrow}{\nabla}_\alpha  \Psi^{i}_\mu  -
m_\Delta \,  \Bar{\Psi}^{i}_\mu   \Psi^{i \mu}  -  g^{\lambda\sigma} \left( \Bar{\Psi}^{i}_\mu
i \gamma^{\mu}{\overset{\leftrightarrow}{\nabla}_\lambda}  \Psi^{i}_\sigma   +  \Bar{\Psi}^{i}_\lambda
i \gamma^{\mu}{\overset{\leftrightarrow}{\nabla}_\sigma}  \Psi^{i}_\mu  \right)  
\nonumber\\
&&+   i  \Bar{\Psi}^{i}_\mu \gamma^\mu \gamma^\alpha\gamma^\nu \overset{\leftrightarrow}{\nabla}_\alpha
\Psi^{i}_\nu + m_\Delta \Bar{\Psi}^{i}_\mu \gamma^\mu \gamma^\nu  \Psi^{i}_\nu +  \frac{g_1}{2}
\,g^{\mu\nu}\bar{\Psi}^i_{\mu} u_\alpha \gamma^\alpha \gamma_5 \Psi^i_{\nu}   
\nonumber \\
&&
+  \frac{g_2}{2} \bar{\Psi}^i_{\mu}  \left( u^\mu  \gamma^\nu + u^\nu \gamma^\mu \right) \gamma_5
\Psi^i_{\nu}   
+
\frac{g_3}{2} \bar{\Psi}^i_{\mu}   u_\alpha  \gamma^\mu \gamma^\alpha \gamma_5  \gamma^\nu \Psi^i_{\nu} 
\biggr\}\,,
\label{gdG}   
\\ 
S_{\Delta \rm N \pi}^{(1,2)} 
&=&
\int d^4 x \sqrt{-g} ~
\Biggl\{  - g_{\pi N \Delta} \bar \Psi \left( g^{\mu \nu} -\gamma^\mu\gamma^\nu\right) u_{\mu,i} \Psi_{\nu,i}  \nonumber\\
&&+
d_3^{(2)} i \bar \Psi f_+^{i \mu \nu} \gamma_5 \gamma_\mu \left( g_{\nu \lambda} -\left[z_n + \frac{1}{2}\right]  \gamma_\nu \gamma_\lambda \right) \Psi^{i \lambda}
+ \text{H.c.} \Biggl\}\,.
\label{gdnG} 
\end{eqnarray}
The $\Delta$ resonances are represented by the Rarita-Schwinger fields
$\Psi^\mu_i$, which contain the isospin-$3/2$ projectors $\xi_{ij}^{\frac{3}{2}}=\delta_{ij}-\tau_{i}
\tau_{j}/3$, i.e.~they satisfy the condition  $\Psi^\mu_i = \xi_{ij}^{\frac{3}{2}} \Psi^\mu_j $.
Further, $g^{\mu\nu}$ is the metric tensor field and $\gamma_\mu \equiv e_\mu^a \gamma_a $, where $ e_\mu^a$ denote
the vielbein gravitational fields. In the photon Lagrangian we
included the mass term ${m_\gamma^2} \, A_\mu A^\mu/2$ to
regularize infrared divergences, and the limit $m_\gamma \to 0$ should be performed  at the end. However,
as it turns out after the calculation, there are actually no IR divergences and this term is thus of no
relevance here.
In Eqs.~(\ref{gdG}) and (\ref{gdnG}), $z_n$ is an off-shell parameter, which we choose
equal to zero in our calculations, 
and we have set the point-transformation parameter $A=-1$
\cite{Tang:1996sq}. The building blocks of the effective Lagrangian
are given as follows:
\begin{eqnarray}
 \overset{\leftrightarrow}{\nabla}_\mu 
 &= &
 \frac{1}{2}( \overset{\rightarrow}{\nabla}_\mu - \overset{\leftarrow}{\nabla}_\mu )
 \,,
 \nonumber\\
 \overset{\rightarrow}{\nabla}_\mu \Psi^i_\nu 
 & = &
 \nabla_\mu^{ij} \Psi^j_\nu = \left[
    \delta^{ij}\partial_\mu + \delta^{ij}\Gamma_\mu-i\delta^{ij} v_\mu^{(s)}-i \epsilon^{ijk}\text{Tr}\left(\tau^k\Gamma_\mu\right) 
    + \frac{i}{2}\delta^{ij}\omega^{ab}_\mu \sigma_{ab}\right]\Psi^j_\nu - \Gamma^{\alpha}_{\mu\nu}\Psi^i_\alpha, 
\nonumber \\ 
\bar\Psi^i_\nu \overset{\leftarrow}{\nabla}_\mu 
& = &
\nabla_\mu^{ij} \Psi^j_\nu =  \bar\Psi^j_\nu \left[
    \delta^{ij}\partial_\mu - \delta^{ij}\Gamma_\mu + i\delta^{ij} v_\mu^{(s)} + i \epsilon^{ijk}\text{Tr}\left(\tau^k\Gamma_\mu\right) 
    - \frac{i}{2}\delta^{ij}\omega^{ab}_\mu \sigma_{ab}\right] - \bar\Psi^i_\alpha\Gamma^{\alpha}_{\mu\nu}, 
\nonumber \\ 
\overset{\rightarrow}{\nabla}_\mu \Psi 
& = & 
\partial_\mu\Psi +\frac{i}{2} \, \omega^{ab}_\mu \sigma_{ab} \Psi + \left( \Gamma_\mu  -i v_\mu^{(s)}\right)\Psi\,,
 \nonumber\\
\bar\Psi \overset{\leftarrow}{\nabla}_\mu
& = &
\partial_\mu\bar\Psi -\frac{i}{2} \, \bar\Psi \, \sigma_{ab} \, \omega^{ab}_\mu - \bar\Psi \left( \Gamma_\mu  -i v_\mu^{(s)}\right) \,, 
\nonumber\\
\omega_\mu^{ab} 
&=&
-\frac{1}{2} \, g^{\nu\lambda} e^a_\lambda \left( \partial_\mu e_\nu^b
- e^b_\sigma \Gamma^\sigma_{\mu \nu} \right),
\nonumber\\
\Gamma^\lambda_{\alpha \beta} &=& \frac{1}{2}\,g^{\lambda\sigma} \left( \partial_\alpha g_{\beta\sigma}
+ \partial_\beta g_{\alpha\sigma} -  \partial_\sigma g_{\alpha\beta} \right)\,, \nonumber\\
 f_+^{ \mu \nu} 
 &=&
 u F^{\mu \nu}_L u^\dagger + u^\dagger F^{\mu \nu}_R u\,, 
 \nonumber\\
F^{\mu \nu}_R &=& 
\partial^\mu r^\nu  - \partial^\nu r^\mu - i [r^\mu,r^\nu]\,,
\nonumber \\
F^{\mu \nu}_L &=& 
\partial^\mu l^\nu  - \partial^\nu l^\mu - i [l^\mu,l^\nu]\,,
\nonumber \\
 f_+^{i \mu \nu} 
 &=& \frac{1}{2} \text{Tr} \left(  f_+^{\mu \nu} \tau^i  \right),
 \nonumber\\
 \chi_+ 
 & = & u^\dagger \chi u^\dagger+u \chi^\dagger u,
 \nonumber\\
\hat \chi_+ 
& = & \chi_+ -\frac{1}{2} \langle \chi_+\rangle,
\nonumber\\
\chi
&=& 2 B_0 (s+ip),
\nonumber\\
u_{\mu,i}
&=& \frac{1}{2}  \text{Tr} \left(  u_{\mu} \tau^i  \right).
\label{Bb1}
\end{eqnarray}
Notice that since the gravitational interaction respects the isospin
symmetry, the amplitude of the $p \to \Delta^+$ transition receives  non-vanishing
contributions only via the  isospin-symmetry breaking effects. 
In Appendix~\ref{sec:A}, the above action is re-written in particle basis and the
corresponding EMT is also specified.

 \section{Gravitational transition form factors to one loop}
\label{II}

Below, we calculate the leading one-loop contributions to the matrix elements of the
EMT for the one-particle states of the delta resonance and the nucleon.
These matrix elements are extracted from the residues of Green's
functions, which have complex poles corresponding to the unstable $\Delta$
states \cite{Gegelia:2010nmt}.
To organize different contributions according to a systematic
expansion we employ the so-called $\epsilon$-counting scheme (also
referred to as the small scale
expansion) \cite{Hemmert:1996xg}\,\footnote{For an alternative power
  counting in ChPT with delta resonances see Ref.~\cite{Pascalutsa:2002pi}.}, 
i.e. the pion lines count as of chiral order $Q^{-2}$, where $Q$ denotes
the soft scale of the order of the pion mass. Further, the nucleon and delta lines
count as $Q^{-1}$,  interaction vertices originating from the effective Lagrangian 
of order $N$ count also as of chiral order $Q^N$, while the vertices
generated by the EMT, which are listed in Appendix \ref{Sec:AppB},
have the orders corresponding 
to the number of the quark mass insertions and derivatives acting on the pion fields. 
Derivatives acting on the nucleon and delta fields count as of chiral order 
$Q^0$. The momentum transfer between the initial and final states counts as of chiral order $Q$, 
therefore in those terms of the EMT which involve full derivatives, these derivatives
also count as chiral order $Q$.  
Integration over loop momenta is counted as chiral order
$Q^4$. Furthermore, the delta-nucleon
mass difference also counts as  order $Q$ within the $\epsilon$-counting scheme.
In diagrams involving electromagnetic radiative corrections, we assign
the chiral order $Q^{-2}$ to the photon line and count the electric
charge $e$ as chiral order $Q$. 
It is understood that the above
described power counting for loop diagrams is realized in the results  
of manifestly Lorentz-invariant calculations only after performing an appropriate
renormalization. We apply the  EOMS scheme of
Refs.~\cite{Gegelia:1999gf,Fuchs:2003qc}.

The matrix element of the total EMT for the transition $p \to \Delta^+ $ can be
parameterized in terms of five form factors as follows \cite{Kim:2022bwn}:
\begin{eqnarray}
&&\langle  \Delta,p_f, s_f | T^{\mu\nu}  | N,p_i, s_i   \rangle 
\nonumber
\\
& = &  \bar u_\alpha(p_f, s_f) \Biggl\{ F_1(t) \left(  g^{\alpha \{\mu}P^{\nu\}} 
+ \frac{m_{\Delta^+}^2-m_p^2 }{\Delta^2} g^{\mu\nu} \Delta^\alpha - \frac{m_{\Delta^+}^2-m_p^2 }{2 \Delta^2} g^{\alpha\{ \mu}\Delta^{\nu\}} - \frac{1}{\Delta^2} P^{\{\mu} \Delta^{\nu\}} \Delta^{\alpha}  \right)  
\nonumber
\\
&+&
F_2(t) \left( P^{\mu}P^{\nu}\Delta^\alpha+ \frac{(m_{\Delta^+}^2-m_p^2 )^2}{4  \Delta^2} g^{\mu\nu} \Delta^\alpha -  \frac{m_{\Delta^+}^2-m_p^2 }{2 \Delta^2} P^{\{\mu} \Delta^{\nu\}} \Delta^{\alpha}  \right)  
\nonumber
\\
&+&
F_3(t) \left( \Delta^\mu  \Delta^\nu - \Delta^2 g^{\mu \nu} \right)  \Delta^\alpha   
\nonumber
\\
&+&
F_4(t) \left( g^{\alpha \{ \mu} \gamma^{\nu\} } +  \frac{2 (m_p + m_{\Delta^+})}{\Delta^2 } g^{\mu\nu} \Delta^\alpha -\frac{m_p + m_{\Delta^+}}{\Delta^2} g^{\alpha\{ \mu}\Delta^{\nu\}} - \frac{1}{\Delta^2}\gamma^{\{\mu} \Delta^{\nu\}} \Delta^{\alpha} \right)
\nonumber
\\
&+&
F_5(t) \left( P^{\{ \mu} \gamma^{\nu\} } \Delta^\alpha  +  \frac{(m_{\Delta^+}^2-m_p^2)  (m_p + m_{\Delta^+})}{ \Delta^2} g^{\mu\nu} \Delta^\alpha - \frac{m_p + m_{\Delta^+}}{ \Delta^2} P^{\{\mu} \Delta^{\nu\}} \Delta^{\alpha}  \right.
\nonumber
\\
&-&\left. \frac{m_{\Delta^+}^2-m_p^2}{2 \Delta^2} \gamma^{\{\mu} \Delta^{\nu\}} \Delta^{\alpha}  \right) \Biggl\} \gamma^5 u(p_i, s_i)\, ,
\end{eqnarray}
where $m_p$ and $m_{\Delta^+}$ are the proton and the $\Delta^+$ masses, respectively, $P = \left( p_f + p_i \right)/2$, $\Delta =  p_f - p_i $ and $t= \Delta^2$.   The curly brackets in the superscripts stand
for symmetrization of the involved indices, e.g., $P^{\{ \mu} \gamma^{\nu\} }=P^{ \mu} \gamma^{\nu }
+P^{\nu} \gamma^{\mu} $.
As mentioned in the introduction, if the isospin symmetry is not broken, the above amplitude is zero.

\subsection{One-loop contributions of the strong interaction to the gravitational transition form factors}

\begin{figure}[htbp]
 \begin{center} 
	\includegraphics[width=0.7\textwidth]{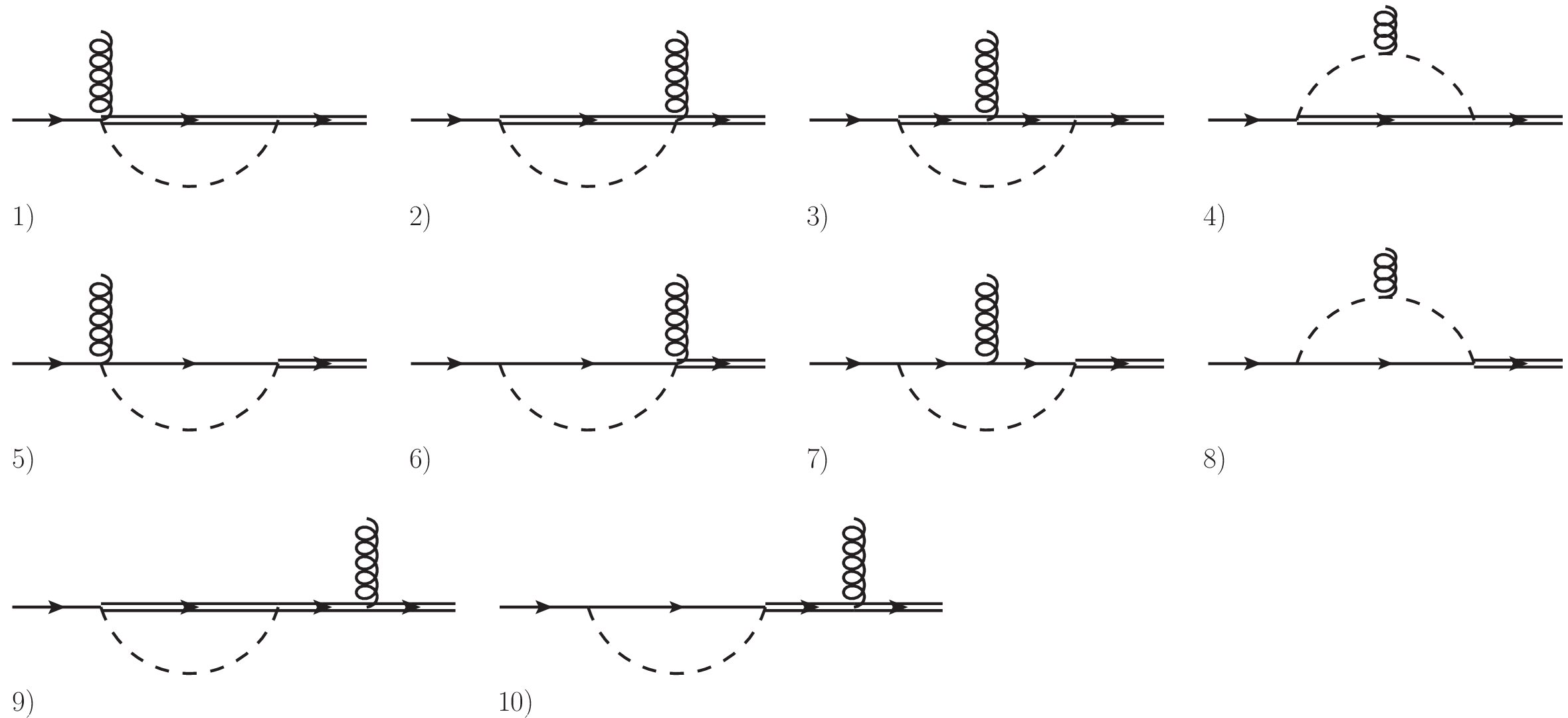}
	\caption{Strong contributions to the gravitational transition
          form factors.~Solid and double lines correspond to nucleons
          and $\Delta$ resonances, respectively. Dashed lines
          represent the pions, while the curly lines correspond to
          gravitons. Initial and final states refer to $p$ and
          $\Delta^+$, respectively, while the baryon lines inside
          loops refer to propagators of one of the following
          particles: $ \{\Delta^{++}, \Delta^+, \Delta^0, p,
          n \} $. Notice that the total contribution of these diagrams vanishes
          in the limit of exact isospin symmetry.
	\label{img:sp}}
      \end{center}    
\end{figure}

To obtain the one-loop contributions to the GTFFs due to strong
isospin-breaking interactions one has to compute 25 diagrams, where there are only 10
topologically differing diagrams and the rest can be obtained by just 
changing the masses and overall factors. These 10 diagrams are depicted in Fig.~\ref{img:sp}.
The isospin symmetry breaking terms of the effective Lagrangian, which
contribute to these one-loop diagrams, are specified in Appendix~\ref{sec:A}. 
We performed the calculations in the particle basis. In the limit of the exact isospin
symmetry, the contributions of the different diagrams exactly cancel each other. Taking
into account the dominant isospin breaking effect, we find that obtained form factors are
proportional to the mass differences within iso-multiplets of nucleons, pions and delta resonances.
The leading contributions are given by terms proportional to the pion mass differences.
This is because these contributions involve integrals, whose integrands are proportional to
\begin{equation} 
\sim \frac{1}{p^2-M_{\pi^+}^2}- \frac{1}{p^2-M_{\pi^0}^2} \simeq  \frac{M_{\pi^+}^2-M_{\pi^0}^2}{\left(p^2-M_{\pi^+}^2\right) \left(p^2-M_{\pi^0}^2\right)}\,,
\label{diffPionPRs}
\end{equation}
where each of the propagators originates from different diagrams that would cancel each other
in the isospin limit. As the mass difference $M_{\pi^+}^2-M_{\pi^0}^2$ counts as of chiral order
two, the right-hand side of Eq.~(\ref{diffPionPRs}) has the same order as each of the terms
in the left-hand side. That is, the total contribution of these diagrams, which is proportional
to the pion mass difference squared, has the same order as the individual diagrams.
On the other hand, the contributions proportional to the proton-neutron mass difference 
are given by integrals, whose integrands are proportional to 
\begin{equation} 
\sim \frac{1}{\slashed p-m_p}- \frac{1}{\slashed p-m_n} \simeq \frac{m_p-m_n}{\left(\slashed p-m_p\right) \left(\slashed p-m_n\right)}\,.
\label{diffNuclPRs}
\end{equation}
As the mass difference $m_p-m_n$ counts as  chiral order two, and the nucleon propagators as
order minus one, the right-hand side of Eq.~(\ref{diffNuclPRs}) has one order higher than
each of the terms on the left-hand side. That is, the total contribution of diagrams is suppressed
by $Q$ relative to the contributions of the individual diagrams.
Analogous power counting holds also for the contributions proportional to the mass
differences of the delta resonances. 
Notice further that isospin breaking vertices other than the mass terms start contributing at
higher orders.  

     Diagrams $1, 2, 4, 5, 6$ and $8$ in Fig.~\ref{img:sp} start contributing at chiral
     order three while the diagrams $3, 7, 9$ and $10$  start
     contributing at chiral order two. This is because the
     leading-order contribution to the
     gravitational-source-baryon-baryon vertex has order zero.\footnote{Actually the
     gravitational-source-baryon-baryon vertex originating from the leading-order Lagrangian
     has two contributions, one of the order zero and the other of the order one. This means
     that diagrams $3, 7, 9$ and $10$ contribute to two different
     chiral orders (2 and 3).~These two contributions cannot be considered separately, because
     otherwise the current will not be conserved.
     This needs to be carefully taken into account when specifying the (possible) power-counting
     violating terms.} 
Thus the diagrams in Fig.~\ref{img:sp} give contributions of orders two and three.
We have verified that the one-loop order result of diagrams in Fig.~\ref{img:sp} does not
contain power counting violating contributions and all ultraviolet divergences can be absorbed
into redefinition of the low-energy coupling constants of the  effective Lagrangian. 
The obtained results for the form factors are too involved to be given as analytic
expressions but are available from the authors upon request in the form of a {\it Mathematica} notebook.
The same applies also to the results of the radiative corrections considered in the next subsection.

\subsection{One-loop radiative corrections to the gravitational transition form factors}
  
\begin{figure}[htbp]
	\centering
	\includegraphics[width=0.7 \textwidth]{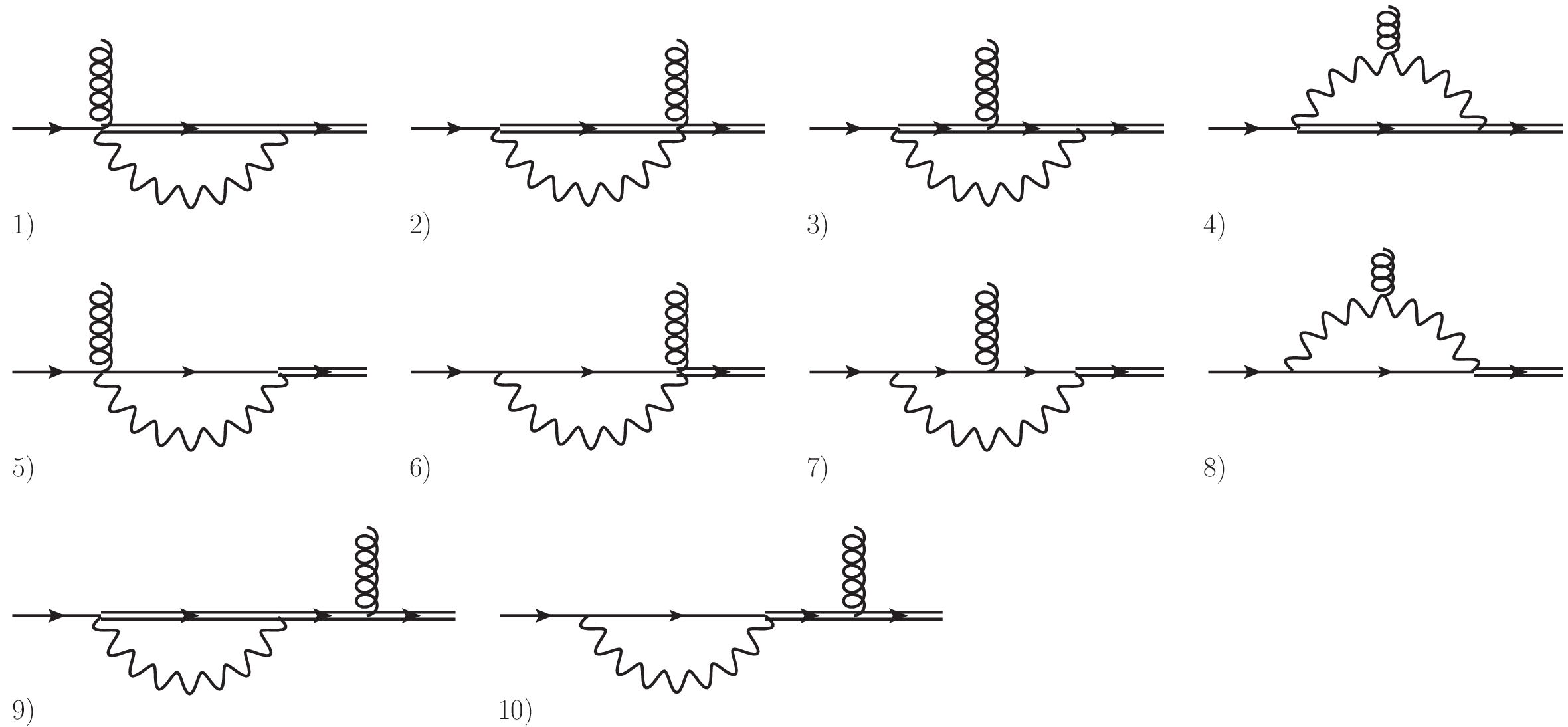}
	\caption{Electromagnetic contributions to the transition form
          factors.~Solid and double solid lines correspond to nucleons
          and $\Delta$ resonances, respectively. Wavy lines denote
          photons, while curly lines represent gravitons.
          	\label{img:em}}
\end{figure}

To obtain the one-loop electromagnetic corrections to the transition
form factors one has to compute the diagrams contributing up to order four, shown in
Fig.~\ref{img:em}. The chiral power counting for Fig.~\ref{img:em} is similar to that for Fig.~\ref{img:sp}.
 In this calculation we do not distinguish between the masses of
 the $\Delta$ states and between the masses of the proton and the
 neutron, i.e. we set $m_{\Delta^{++}} = m_{\Delta^{+}} =
 m_{\Delta^{0}} = m_{\Delta^{-}}$ and $m_p = m_n$.

Analogously to the strong-interaction contributions, we found that the
one-loop order result of diagrams shown in Fig.~\ref{img:em} does not
involve power-counting violating terms, and all ultraviolet
divergences can be absorbed into redefinition of the low-energy coupling constants of
the most general effective Lagrangian.

\subsection{Numerical results for the gravitational transition form factors}

\begin{figure}[htbp]
	\centering
	\includegraphics[width=0.75 \textwidth]{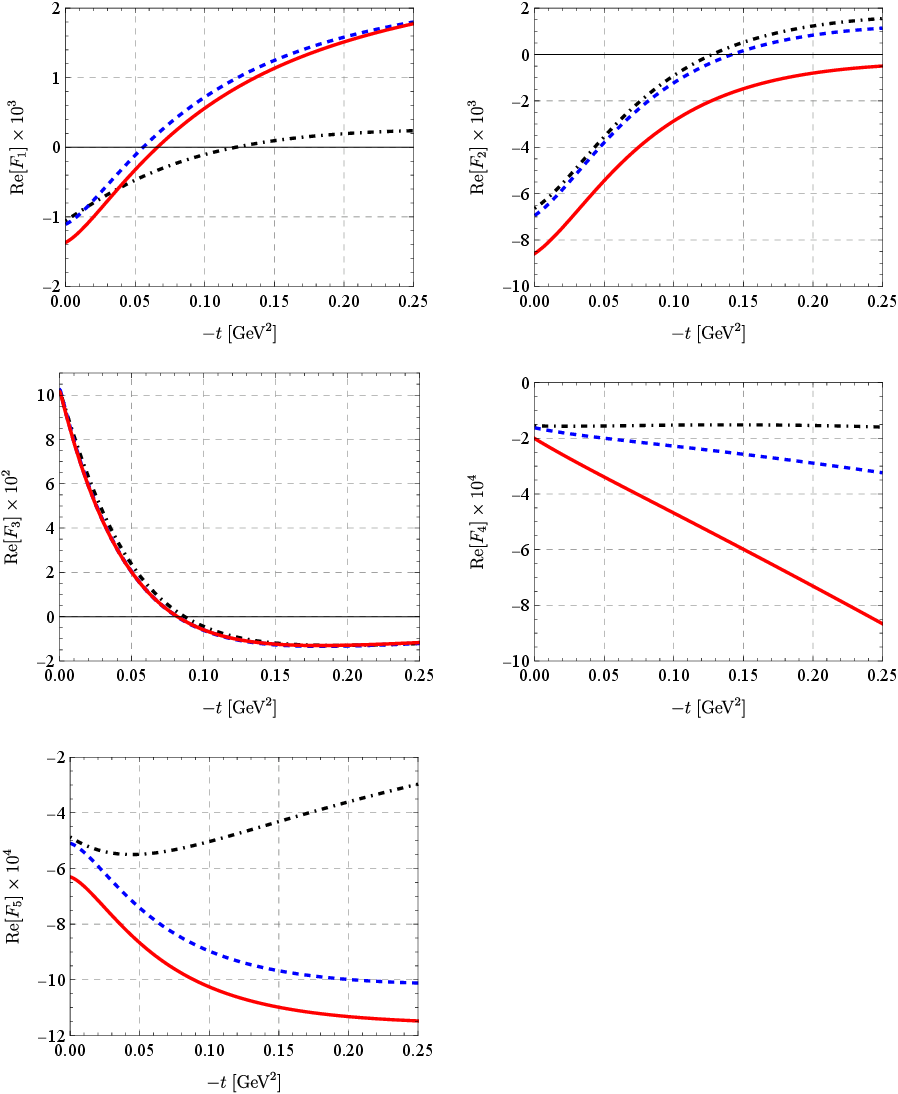}
	\caption{The real parts of the $p \to \Delta^+ $ transition
          form factors.~Dash-dotted (black),  dashed (blue) and solid 
          (red) lines correspond to the form factors containing
          contributions of loop diagrams with inner pion and nucleon lines
          only, diagrams with inner pion and nucleon lines plus radiative
          corrections, and all loop contributions,
          respectively.  }
	\label{imgR:em}
\end{figure}

\begin{figure}[htbp]
	\centering
	\includegraphics[width=0.75 \textwidth]{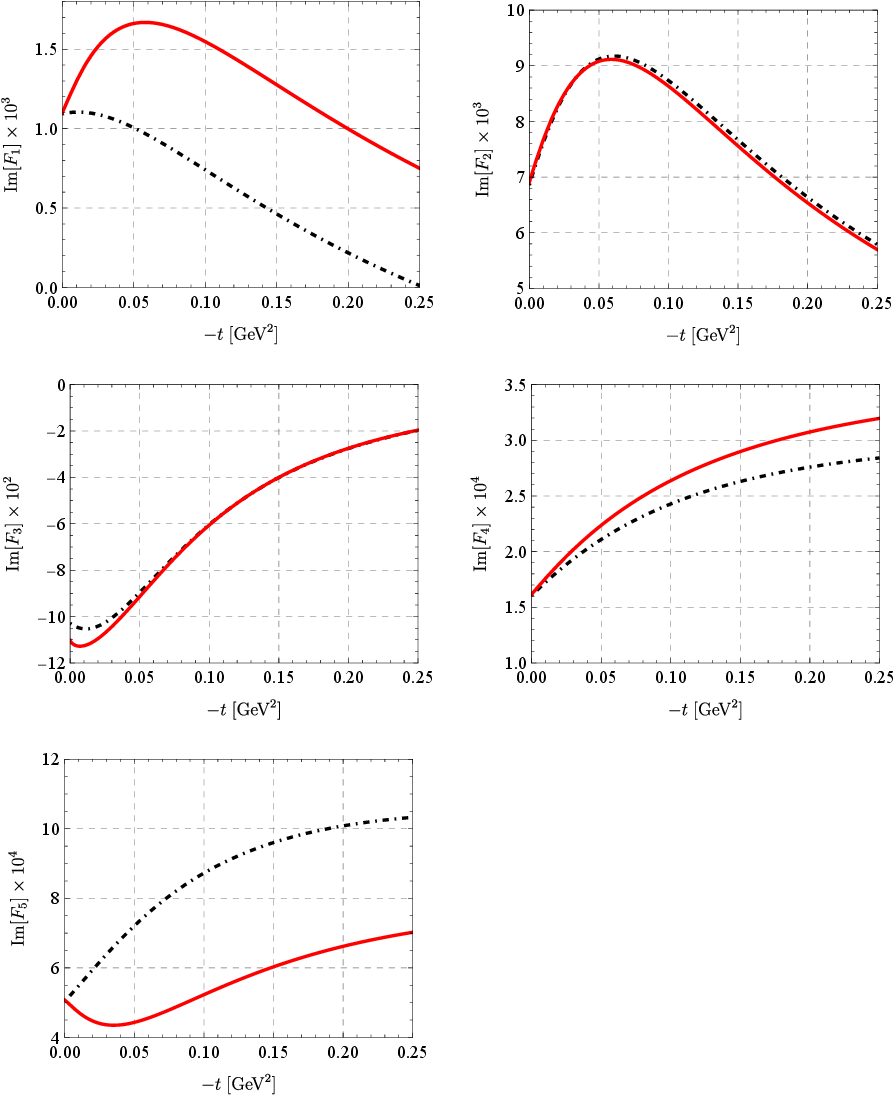}
	\caption{Imaginary parts of the $p \to \Delta^+ $ transition form factors.~Dash-dotted (black), and  solid (red) lines correspond to the form factors containing contributions of loop diagrams with inner pion and nucleon lines only, and diagrams with inner pion and nucleon lines plus radiative corrections, respectively. }
	\label{imgI:em}
\end{figure}

In Figs.~\ref{imgR:em} and \ref{imgI:em}, we present the numerical results of the obtained
strong and electromagnetic contributions to the real and imaginary parts of the transition form
factors, respectively.  
Notice here that the imaginary parts of the calculated form factors are generated solely
by the loop contributions with internal nucleon lines. 
For the numerical results, we used the following values of the involved
parameters:
\begin{eqnarray} 
 g_A & = & 1.289, \quad
g = 1.35,  \quad
m_{\pi^0} = 0.135,  \quad
m_p = 0.938,  \quad
m_n = 0.940,  \quad \nonumber\\
m_\Delta & =  &1.232,  \quad
F = 0.092,  \quad
m_{\pi^+}  =  0.140,  \quad
m_{\Delta^{++}} = 1.231 ,  \quad
m_{\Delta^+} = m_{\Delta},   \nonumber\\
m_{\Delta^0} & = & 1.233,  \quad
g_1 = 9 g_A/5, \quad 
e = 0.303 \,, \quad d_3^{(2)}=2.72\, {\rm GeV}^{-1}\,, 
\label{parameters}
\end{eqnarray}
where the various masses and the pion decay constant $F$ are given in GeV. We used
the $SU(6)$ symmetry estimation for the coupling constants $g_1$,
$g_A$ and $g$ taken from Ref.~\cite{Siemens:2016hdi},  for the masses of delta resonances
we used estimations of Refs.~\cite{Epelbaum:2007sq,Epelbaum:2008td}, 
$d_3^{(2)}$ corresponds to $b_1/2$ of Ref.~\cite{Bernard:2012hb}, 
 while the remaining values have
been taken from the PDG \cite{ParticleDataGroup:2020ssz}. 

The plots demonstrate that the diagrams with radiative corrections give smaller
contributions than the ones with pion loops in line with the power counting estimations.  
On the other hand the groups of diagrams with internal nucleon and delta lines give
comparable contributions.

\section{Conclusions and outlook} 
\label{summary}

In the framework of manifestly Lorentz-invariant ChPT for pions, nucleons, photons and
the delta resonances interacting with an external gravitational field, we calculated the leading
one-loop contributions to the matrix element of the EMT corresponding to the $p \to \Delta^+ $
transition and extracted the resulting gravitational transition form factors. 
	As the gravitational interaction respects the isospin
        symmetry, the amplitude of the $p \to \Delta^+ $ transition
        receives non-vanishing contributions due to isospin symmetry breaking. The results of
        the current work take into account the leading-order electromagnetic and strong
        isospin-breaking effects. 	
	Ultraviolet divergences and power counting violating pieces
        generated by loop diagrams in the manifestly Lorentz-invariant
        formulation of ChPT can be treated using the EOMS renormalization scheme of
        Refs.~\cite{Gegelia:1999gf,Fuchs:2003qc}.
However, at the order of our calculations, the one-loop contributions
to the form factors are found to be free of contributions that violate
the chiral power counting. This is consistent with the absence of
tree-level contributions at the considered order. For this reason, our
results involve no free parameters and can be regarded as 
predictions of ChPT. Notice, however, that the empirical information
on the mass splittings between the $\Delta$ resonance states, which enters as an
input in our calculations, is presently rather poor. 
Numerical results for the obtained transition form factors demonstrate
that the electromagnetic and strong isospin violating effects give
contributions of comparable sizes. This holds true for contributions
with both internal nucleon and delta lines.

\acknowledgments 

This work was supported in part by 
DFG and NSFC through funds provided to the Sino-German CRC 110
“Symmetries and the Emergence of Structure in QCD” (NSFC Grant
No. 11621131001, DFG Project-ID 196253076 - TRR 110),
by CAS through a President’s International Fellowship Initiative (PIFI)
(Grant No. 2018DM0034), by the VolkswagenStiftung
(Grant No. 93562), by the MKW NRW under the funding code NW21-024-A, by the EU Horizon 2020 research and
innovation programme (STRONG-2020, grant agreement No. 824093), by Guangdong Provincial funding with Grant
No. 2019QN01X172, the National Natural Science Foundation of China
with Grant No. 12035007 and No. 11947228, Guangdong Major Project of
Basic and Applied Basic Research No. 2020B0301030008, and the Department of Science and
Technology of Guangdong Province with Grant No. 2022A0505030010.

\appendix

\section{Isospin-symmetry breaking terms } 
\label{sec:A}
To obtain the leading isospin breaking effects due to the strong interaction we
distinguish between the masses of the delta resonances, and also between the masses of the proton and
the neutron, and the charged and neutral pions.
We rewrite the action using the physical basis, instead of the isospin basis,
by writing the fields explicitly as follows:
\begin{eqnarray}
\Psi &=& \begin{pmatrix}
 \Psi_p  \\ \Psi_n \end{pmatrix}, \nonumber\\ 
 \Psi_{\mu,1} &=& \frac{1}{\sqrt{2}}
 \begin{pmatrix}
 \frac{1}{\sqrt{3}} \Delta^0_\mu - \Delta^{++}_\mu \\ \Delta^-_\mu  -  \frac{1}{\sqrt{3}} \Delta^+_\mu 
\end{pmatrix} , ~ 
 \Psi_{\mu,2} = -\frac{i}{\sqrt{2}}
 \begin{pmatrix}
 \frac{1}{\sqrt{3}} \Delta^0_\mu + \Delta^{++}_\mu \\ \Delta^-_\mu +  \frac{1}{\sqrt{3}} \Delta^+_\mu 
\end{pmatrix}, ~ 
 \Psi_{\mu,3} = \sqrt{\frac{2}{3}}
 \begin{pmatrix}
\Delta^+_\mu \\ \Delta^0_\mu 
\end{pmatrix}\,, \nonumber
\\
\pi^1& =& \frac{1}{\sqrt{2}} \left( \pi^+ + \pi^- \right),\, \pi^2 = \frac{i}{\sqrt{2}} \left( \pi^+ -\pi^- \right),\, \pi^3 =  \pi^0.
\end{eqnarray} 
We substitute the above definition of the fields into Eqs.~(\ref{PionAction}), (\ref{PiNAction}), (\ref{gdG}) and (\ref{gdnG}). The terms relevant for the leading one-loop order contributions to the $p \to \Delta^+ $
transition are given by
\begin{eqnarray}
S_{\rm \pi}^{(2)} 
&=& \int d^4x \sqrt{-g}\,  \left\{  \frac{1}{2} \partial_\mu \pi^0 \partial^\mu \pi^0 - \frac{1}{2} M_0^2 \pi^0\pi^0 +  \partial_\mu \pi^+ \partial^\mu \pi^- - M_{\pi^+}^2 \pi^+\pi^-\ \right\}\,,
\\
S_{\rm N \pi}^{(1)}  
& = & 
\int d^4x \sqrt{-g}\,  \left\{\, \bar\Psi_p \, i  \gamma^\mu \overset{\leftrightarrow}{\nabla}_\mu \Psi_p  - m_p \bar\Psi_p\Psi_p + \bar\Psi_n \, i  \gamma^\mu \overset{\leftrightarrow}{\nabla}_\mu \Psi_n -m_n \bar\Psi_n\Psi_n \right.
\nonumber
\\
&+& \left.
\frac{g_A}{2F} \left( \partial_\mu\pi^0 \left[ \bar\Psi_n \gamma^\mu \gamma^5 \Psi_n - \bar\Psi_p \gamma^\mu \gamma^5 \Psi_p\right] - \sqrt{2} \left[ \partial_\mu \pi^- \bar\Psi_n \gamma^\mu \gamma^5 \Psi_p +\partial_\mu \pi^+ \bar\Psi_p \gamma^\mu \gamma^5 \Psi_n   \right] \right)\right\} \,,
\nonumber
\\
\\ 
S_{\Delta \pi}^{(1)} 
& = &  - \int d^4 x  \sqrt{-g} \biggl\{  
\biggl[ \sum_{i\in \{++,+,0,- \}} \Bar{\Delta}^{i \mu}  \,  i \gamma^\alpha \overset{\leftrightarrow}{\nabla}_\alpha  \Delta^{i}_\mu  -
m_{\Delta^i} \,  \Bar{\Delta}^{i}_\mu   \Delta^{i \mu} 
\nonumber\\
&&
-  g^{\lambda\sigma} \left( \Bar{\Delta}^{i}_\mu
i \gamma^{\mu}{\overset{\leftrightarrow}{\nabla}_\lambda}  \Delta^{i}_\sigma   +  \Bar{\Delta}^{i}_\lambda
i \gamma^{\mu}{\overset{\leftrightarrow}{\nabla}_\sigma}  \Delta^{i}_\mu  \right)  
+  i  \Bar{\Delta}^{i}_\mu \gamma^\mu \gamma^\alpha\gamma^\nu \overset{\leftrightarrow}{\nabla}_\alpha
\Delta^{i}_\nu + m_{\Delta^{i}} \Bar{\Delta}^{i}_\mu \gamma^\mu \gamma^\nu  \Delta^{i}_\nu \biggr] 
\nonumber\\
&&
+ \frac{1}{6 F}  \left[\bar \Delta^0_\mu O_1^{\mu \nu \alpha} \Delta^0_\nu  \partial_\alpha \pi^0 - \sqrt{6} \bar \Delta^-_\mu O_1^{\mu \nu \alpha}  \Delta^0_\nu  \partial_\alpha \pi^-  -2 \sqrt{2} \bar \Delta^+_\mu O_1^{\mu \nu \alpha} \Delta^0_\nu  \partial_\alpha \pi^+ \right.
\nonumber\\
&& - \sqrt{6} \bar \Delta^0_\mu O_1^{\mu \nu \alpha}  \Delta^-_\nu  \partial_\alpha \pi^+ - 2\sqrt{2} \bar \Delta^0_\mu O_1^{\mu \nu \alpha}  \Delta^+_\nu  \partial_\alpha \pi^- +  3 \bar \Delta^-_\mu O_1^{\mu \nu \alpha}  \Delta^-_\nu  \partial_\alpha \pi^0 
\nonumber\\
&& -
\bar \Delta^+_\mu O_1^{\mu \nu \alpha}  \Delta^+_\nu  \partial_\alpha \pi^0- \sqrt{6} \bar \Delta^+_\mu O_1^{\mu \nu \alpha}  \Delta^{++}_\nu  \partial_\alpha \pi^- - \sqrt{6} \bar \Delta^{++}_\mu O_1^{\mu \nu \alpha}  \Delta^{+}_\nu  \partial_\alpha \pi^+ 
\nonumber\\
&& 
\left. -3 \bar \Delta^{++}_\mu O_1^{\mu \nu \alpha}  \Delta^{++}_\nu  \partial_\alpha \pi^0
\right]\Biggl\}\,,
\\
S_{\Delta \rm N \pi}^{(1)} 
&=& 
\int d^4x \sqrt{-g}\,   \frac{g_{\pi n \Delta}}{F} \Biggl\{ \bar \Psi_n \partial_\mu  \pi^+ O_2^{\mu \nu} \Delta^-_\nu  - \bar \Psi_p \partial_\mu  \pi^- O_2^{\mu \nu} \Delta^{++}_\nu  + \frac{1}{\sqrt{3}} \left(\sqrt{2} ~\bar \Psi_n \partial_\mu \pi^0 O_2^{\mu \nu} \Delta^0_\nu   \right.
\nonumber
\\
&&
 - \bar \Psi_n \partial_\mu \pi^- O_2^{\mu \nu} \Delta^+_\nu
+ \left. \bar \Psi_p \partial_\mu \pi^+ O_2^{\mu \nu} \Delta^0_\nu +  \sqrt{2}~ \bar \Psi_p \partial_\mu \pi^0 O_2^{\mu \nu} \Delta^+_\nu \right) \Biggl\}\,,
\end{eqnarray}
where $O_1^{\mu \nu \alpha} = g_1 \gamma^\alpha\gamma^5 g^{\mu\nu} + g_2 (g^{\mu \alpha}\gamma^\nu\gamma^5
+ g^{\nu \alpha}\gamma^\mu \gamma^5) + g_3 \gamma^\mu \gamma^\alpha \gamma^5 \gamma^\nu$ and
$  O_2^{\mu \nu} = g^{\mu \nu} - \gamma^\mu \gamma^\nu$. To arrive at these results, we expanded the matrix
$u$ of pion fields and kept only the first nontrivial term, i.e. $u= 1 + i/(2 F) \tau^i \pi^i
+ \mathcal{O}(1/F^2).$

The mass splittings within iso-multiplets are not just due to strong isospin breaking but also
receive important contributions from the electromagnetic interaction. However, there
is no point at separating these contributions here, and such a separation is anyway
afflicted with some uncertainties, see e.g. the pedagogical discussion in~Ref.~\cite{Meissner:2022cbi}.

To obtain the leading isospin breaking effects due to the diagrams with radiative corrections
(i.e. with photon propagators) we do not distinguish between the masses of the isospin partners,
i.e. we take $m_{\Delta^{++}} = m_{\Delta^{+}} = m_{\Delta^{0}} = m_{\Delta^{-}}$, and $m_p = m_n$.
For the external sources, we take the following expressions:
\begin{eqnarray}
r_\mu 
&=&
 l_\mu = - e ~A_\mu \frac{\tau_3}{2},
\\
v_\mu^{s}  
&=&
- \frac{e}{2} ~A_\mu,
\end{eqnarray}
where $e$ is the electric charge of the proton.

\section{The energy-momentum tensor}
\label{Sec:AppB}
     
Using the definition of the EMT for bosonic matter fields interacting with the gravitational metric field,
\begin{eqnarray}
T_{\mu\nu} (g,\psi) & = & \frac{2}{\sqrt{-g}}\frac{\delta S_{\rm m} }{\delta g^{\mu\nu}}\,,
\label{EMTMatter}
\end{eqnarray}
we obtain in flat spacetime from the action terms of Eqs.~(\ref{Gg}) and (\ref{PionAction}):
\begin{eqnarray}
T^{{(2)}}_{\gamma , \mu\nu} 
&=&
F_{\mu}^\alpha F_{\alpha \nu}+ m_\gamma^2 A_\mu A_\nu+\eta_{\mu \nu} \left( \frac{1}{4} F_{\alpha \beta }F^{\alpha \beta } -\frac{m_\gamma^2}{2} A_\alpha A^\alpha  \right),\,
\label{PhotonEMT}
\\
T^{{(2)}}_{\pi , \mu\nu}  & = &  \frac {F^2}{4}\, {\rm Tr} ( D_\mu U  (D_\nu U)^\dagger)
-\frac{ \eta_{\mu\nu}}{2} \left\{ \frac {F^2}{4}\, {\rm Tr} ( D^\alpha U  (D_\alpha U)^\dagger )
+  \frac{F^2}{4}\,{\rm Tr}(\chi U^\dagger +U \chi^\dagger) \right\} 
\nonumber \\
&& + \left( \mu \leftrightarrow
\nu \right),
\label{PionEMT}
\end{eqnarray}
where $\eta_{\mu\nu}$ is the Minkowski metric tensor with the
signature $(+,-,-,-)$.
For the fermionic fields interacting with the gravitational vielbein fields we use the definition
\cite{Birrell:1982ix} 
\begin{eqnarray}
T_{\mu\nu}  (g,\psi) & = & \frac{1}{2 e} \left[ \frac{\delta S }{\delta e^{a \mu}} \,e^{a}_\nu
+ \frac{\delta S }{\delta e^{a \nu}} \,e^{a}_\mu  \right]\,.
\label{EMTfermion}
\end{eqnarray}
The action of Eq.~(\ref{PiNAction}) leads to the following expression for the EMT  in
flat spacetime:
\begin{eqnarray}
T^{{  (1)}}_{\rm N, \mu\nu}  & = &  \frac{i}{2} \bar\Psi \,  \gamma_\mu  \overset{\leftrightarrow}{D}_\nu \Psi - \frac{\eta_{\mu\nu}}{2}\biggl(\, \bar\Psi \, i  \gamma^\alpha \overset{\leftrightarrow}{D}_\alpha \Psi -m \bar\Psi\Psi  \biggr) + \left( \mu \leftrightarrow \nu \right)   \,,
\label{MEMT}
\end{eqnarray}
while Eqs.~(\ref{gdG}) and (\ref{gdnG}) 
lead to the following expressions:
\begin{eqnarray}
  T^{{  (1)}}_{\Delta \pi  ,\mu\nu} & = &   -\Bar{\Psi}^{i}_\mu  \,  i  \gamma^\alpha \overset{\leftrightarrow}{D}_\alpha  \Psi^{i}_\nu + \Bar{\Psi}^{i}_\alpha  \,  i  \gamma^\alpha \overset{\leftrightarrow}{D}_\mu  \Psi^{i}_\nu   +
       \Bar{\Psi}^{i}_\mu  \,  i  \gamma^\alpha \overset{\leftrightarrow}{D}_\nu  \Psi^{i}_\alpha + m_\Delta \Bar{\Psi}^{i}_\mu   \Psi^{i}_\nu - \frac{i}{2} \, \bar\Psi^i_\alpha \,  \gamma_\mu  \overset{\leftrightarrow}{D}_\nu \Psi^{i\alpha}
\nonumber\\
&& +\frac{i}{2}   \biggl(  \bar\Psi^i_\mu \,  \gamma_\nu  \overset{\leftrightarrow}{D}_\alpha \Psi^{i\alpha} + \bar\Psi^{i\alpha} \,  \gamma_\nu \overset{\leftrightarrow}{D}_\alpha \Psi^{i}_\mu   -  \bar\Psi^i_\mu \,  \gamma_\nu \gamma^\alpha\gamma_\beta  \overset{\leftrightarrow}{D}_\alpha \Psi^{i,\beta}  - \bar\Psi^i_\alpha \gamma^\alpha \gamma_\nu \gamma^\beta  \overset{\leftrightarrow}{D}_\mu \Psi^{i}_\beta
\nonumber\\
&& 
- \bar\Psi^i_\alpha \gamma^\alpha\gamma^\beta \gamma_\nu  \overset{\leftrightarrow}{D}_\beta \Psi^{i}_\mu  \biggl)
+\frac{i}{4}\,\partial^\lambda \biggl[\bar\Psi^{i, \alpha}  \biggl( \gamma_\mu \eta_{\lambda [\alpha}\eta_{\beta] \mu} +\eta_{\lambda \mu}\eta_{\nu [\alpha}\gamma_{\beta]} + \eta_{\mu\nu} \eta_{\lambda[\beta}\gamma_{\alpha]} \biggr)
\Psi^{i,\beta}  \biggr]
\nonumber\\
&&  
 - \frac{m_\Delta}{2} \, \left( \bar\Psi^i_\mu \,  \gamma_\nu \gamma^\alpha \Psi^{i}_\alpha + \bar\Psi^i_\alpha \,  \gamma^\alpha \gamma_\nu \Psi^{i}_\mu \right)  
-\frac{g_1}{4} \left[ 2 \bar{\Psi}^i_{\mu} u_\alpha \gamma^\alpha \gamma_5 \Psi^i_{\nu} +  \bar{\Psi}^{i,\alpha}  u_\mu  \gamma_\nu \gamma_5 \Psi^i_{\alpha}  \right]
\nonumber\\
&&  
-\frac{g_2}{4} \left[  2 \bar{\Psi}^i_\mu u_\nu  \gamma^\alpha \gamma_5 \Psi^i_{\alpha}  + 2 \bar{\Psi}^i_\alpha u_\nu  \gamma^\alpha \gamma_5 \Psi^i_{\mu}   + \bar{\Psi}^{i,\alpha} u_\alpha  \gamma_\nu \gamma_5 \Psi^i_{\mu} + \bar{\Psi}^{i}_\mu u_\alpha  \gamma_\nu \gamma_5 \Psi^{i \alpha} \right]   \nonumber \\ 
    &&  - \frac{g_3}{4} \left[ \bar{\Psi}^i_{\mu} u_\alpha \gamma_\nu \gamma^\alpha \gamma_5  \gamma^\beta \Psi^i_{\beta} 
      +\bar{\Psi}^i_{\beta} u_\alpha  \gamma^\beta \gamma^\alpha \gamma_5  \gamma_\nu \Psi^i_{\mu} + \bar{\Psi}^i_{\alpha} u_\mu  \gamma^\alpha \gamma_\nu \gamma_5  \gamma^\beta \Psi^i_{\beta} \right]
      \nonumber \\ && 
+ \frac{\eta_{\mu\nu} }{2} \biggl[  
     \, \Bar{\Psi}^{i}_\alpha  \,  i \gamma^\beta \overset{\leftrightarrow}{D}_\beta  \Psi^{i\alpha } -  m_\Delta \, \Bar{\Psi}^{i}_\alpha   \Psi^{i\alpha}  -\Bar{\Psi}^{i}_\alpha  i \gamma^{\alpha}{\overset{\leftrightarrow}{D}_\beta}  \Psi^{i\beta} -\Bar{\Psi}^{i\alpha}i \gamma^{\beta}{\overset{\leftrightarrow}{D}_\alpha}  \Psi^{i}_\beta 
 \nonumber\\ &&  
 +i  \Bar{\Psi}^{i}_\rho \gamma^\rho \gamma^\alpha\gamma^\lambda \overset{\leftrightarrow}{D}_\alpha \Psi^{i}_\lambda + m_\Delta \Bar{\Psi}^{i}_\alpha \gamma^\alpha \gamma^\beta  \Psi^{i}_\beta  + 
   \frac{g_1}{2} \, \bar{\Psi}^i_{\beta} u_\alpha  \gamma^\alpha \gamma_5 \Psi^{i\beta}   
   \nonumber \\
   &&   
   + \frac{g_2}{2} \bar{\Psi}^{i\alpha}   \left( u_\alpha  \gamma_\beta +u_\beta   \gamma_\alpha  \right) \gamma_5 \Psi^{i\beta}  
   + \frac{g_3}{2} \bar{\Psi}^i_{\alpha}   u_\beta  \gamma^\alpha \gamma^\beta \gamma_5  \gamma^\lambda \Psi^i_{\lambda} \biggr] 
   +  \left( \mu \leftrightarrow \nu \right)  \,,
   \label{EMTB}
   \end{eqnarray}
\begin{eqnarray}
T^{(1,2)}_{\pi N\Delta, \mu\nu}  &=&g_{\pi N\Delta} \biggl\{  \frac{1}{2} ~\eta_{\mu \nu} \left[ \bar \Psi^i_\alpha u^{\alpha}_i\Psi + \bar \Psi u^{\alpha}_i   \Psi^i_\alpha - \bar \Psi^i_\alpha \gamma^\alpha \gamma^\beta u_{\beta }^i \Psi - \bar \Psi  \gamma^\beta  \gamma^\alpha u_{\beta }^i  \Psi^i_\alpha \right] -  \bar\Psi^i_\mu u_{\nu}^i\Psi \nonumber
\\
&&  -  \bar\Psi u_{\nu}^i \Psi^i_\mu +\frac{1}{2}  ~ \left[ \bar \Psi^i_\mu \gamma_\nu  \gamma^\alpha u_{\alpha }^i \Psi + \bar \Psi^i_\alpha \gamma^\alpha  \gamma_\mu u_{\nu}^i \Psi + \bar \Psi  \gamma^\alpha \gamma_\nu  u_{\alpha }^i  \Psi^i_\mu + \bar \Psi  \gamma_\mu  \gamma^\alpha u_{\nu}^i  \Psi^i_\alpha \right] \biggl\}  \nonumber
\\
&& +\frac{i }{2} d_3^{(2)} \biggl\{\bar \Psi f_{+, \mu \beta }^{i } \gamma_5 \gamma_\nu \tilde{\Psi}^{i \beta} + 2 \bar \Psi f_{+, \alpha \mu }^{i } \gamma_5 \gamma^\alpha \Psi^{i}_\nu   -  \eta_{\mu \nu}  \bar \Psi f_{+, \alpha \beta }^{i } \gamma_5 \gamma^\alpha \tilde{\Psi}^{i \beta}  
\nonumber
\\
&&  - \left[z_n+ \frac{1}{2} \right] \left(\bar \Psi f_{+, \alpha \mu }^{i} \gamma_5 \gamma^\alpha \gamma_\nu \gamma^\beta \Psi^{i}_\beta   + \bar \Psi f_{+, \alpha \beta }^{i} \gamma_5 \gamma^\alpha \gamma^\beta \gamma_\mu \Psi^{i}_\nu   \right) \biggl\} + \left( \mu \leftrightarrow \nu \right)\, ,
\end{eqnarray}
where the covariant derivatives ${D}$ acting on spin-1/2 and spin-3/2 fields coincide with $\nabla$ of Eq.~(\ref{Bb1}) with 
$ \Gamma_{\mu\nu}^\beta = \omega_{\mu}^{ab} = 0$. The superscripts in the expressions of EMT indicate the orders which are assigned to the corresponding terms of the action (effective Lagrangian).




\end{document}